\documentclass[12pt]{article}
\usepackage[dvips]{epsfig}
\begin{document}
\begin{flushright}
hep-th/9810093
\end{flushright}
\vspace{20mm}
\begin{center}
{\LARGE How `Complex' is the Dirac Equation? }
\\
\vspace{20mm}
{\bf Francesco Antonuccio \\}
\vspace{4mm}
Department of Physics,\\ The Ohio State University,\\ Columbus, OH 43210, USA\\
\vspace{4mm}
\end{center}
\vspace{10mm}

\begin{abstract}
A representation of the Lorentz group
is given in terms of $4 \times 4$ matrices defined over a simple
non-division algebra. The transformation properties of
the corresponding four component spinor are studied, and 
shown to be equivalent to the transformation properties of
the usual complex Dirac spinor. 
As an application, we show that there exists
an algebra of automorphisms of the complex
Dirac spinor that leave the transformation
properties of its eight real components 
invariant under any given Lorentz transformation.   
Interestingly, the representation of the Lorentz
group presented here has a natural embedding in
a cover of SO(3,3) instead of the conformal symmetry SO(2,4).

\end{abstract}
\newpage

\baselineskip .25in

\section{Introduction}
This article is motivated by the simple observation that the 
transformation properties of the eight real components
of a complex Dirac spinor under a Lorentz
transformation may be concisely 
formulated without any explicit reference
to complex-valued quantities.
This is accomplished by a representation of the 
Lorentz group using $4 \times 4$ matrices
defined over a simple non-division algebra.
After studying how this new representation is related to
the usual complex one, we establish
an automorphism symmetry of the complex Dirac spinor.
We also discuss natural embeddings of this new representation
into a maximal group, which turns out to be SO(3,3),
and thus not equal to the usual conformal symmetry SO(2,4).

To begin, we revisit the familiar Lie algebra of the
Lorentz group O(1,3). 
\section{The Lorentz Algebra}
\subsection{A Complex Representation}
Under Lorentz transformations, the complex Dirac 4-spinor $\Psi_{\bf C}$
transforms as follows \cite{ryder}:
\begin{equation}
\Psi_{\bf C} \rightarrow \left(
\begin{array}{cc}
       e^{\frac{{\rm i}}{2} 
   \mbox{${\bf \sigma \cdot}$} 
          ( \mbox{${\bf \theta}$} - {\rm i}
           \mbox{${\bf \phi}$})}
            & 0 \\
       0 &  e^{\frac{{\rm i}}{2} 
   \mbox{${\bf \sigma \cdot}$} 
          ( \mbox{${\bf \theta}$} + {\rm i}
           \mbox{${\bf \phi}$})} 
\end{array}
\right) \cdot \Psi_{\bf C},
\label{lorentz1} 
\end{equation}
where ${\bf \sigma} = (\sigma_{x},\sigma_{y},\sigma_{z})$ represents
the well known Pauli spin matrices:
\begin{equation}
\sigma_x = \left(
\begin{array}{cc}
 0 & 1 \\
 1 & 0
\end{array}
\right), \hspace{5mm}
\sigma_y = \left(
\begin{array}{cc}
 0 & -{\rm i} \\
 {\rm i} & 0
\end{array}
\right), \hspace{5mm}
\sigma_z = \left(
\begin{array}{cc}
 1 & 0 \\
 0 & -1
\end{array}
\right).
\end{equation}
The three real parameters ${\bf \theta} = (\theta_1,\theta_2,\theta_3)$
correspond to the generators for spatial rotations,
while ${\bf \phi} = (\phi_1,\phi_2,\phi_3)$ represents 
Lorentz boosts along each of the coordinate axes.
There are thus six real numbers parameterizing a given element
in the Lorentz group.

Let us now introduce the six matrices $E_i$ and $F_i$, $i=1,2,3$,
by writing
\begin{equation}
\begin{array}{ccc}
E_1 = \frac{1}{2} \left( 
\begin{array}{cc}
\sigma_x & 0 \\
0 & -\sigma_x
\end{array}
\right)
& 
E_2 = -\frac{{\rm i}}{2} \left( 
\begin{array}{cc}
\sigma_y & 0 \\
0 & \sigma_y
\end{array}
\right)
 & E_3 = \frac{1}{2} \left( 
\begin{array}{cc}
\sigma_z & 0 \\
0 & -\sigma_z
\end{array}
\right)  \\
F_1 = \frac{{\rm i}}{2} \left( 
\begin{array}{cc}
\sigma_x & 0 \\
0 & \sigma_x
\end{array}
\right)
 & 
F_2 = \frac{1}{2} \left( 
\begin{array}{cc}
\sigma_y & 0 \\
0 & -\sigma_y
\end{array}
\right)
 & 
F_3 = \frac{{\rm i}}{2} \left( 
\begin{array}{cc}
\sigma_z & 0 \\
0 & \sigma_z
\end{array}
\right).
\end{array}
\label{ef}
\end{equation}
Then the Lorentz transformation (\ref{lorentz1}) may be written as follows:
\begin{equation}
\Psi_{\bf C} \rightarrow \exp{(\phi_1 E_1 - \theta_2 E_2 + \phi_3 E_3 + 
\theta_1 F_1 + \phi_2 F_2 + \theta_3 F_3)} \cdot  \Psi_{\bf C}.
\label{transformD}
\end{equation}
It is a straightforward exercise to check that the matrices
$E_i$ and $F_i$ defined in (\ref{ef}) satisfy the following
commutation relations:
\begin{equation}
\begin{array}{llll}
[E_1,E_2] = E_3 & [F_1,F_2] = -E_3 & [E_1,F_2] = F_3 & [F_1,E_2]=F_3 \\
\mbox{}[E_2,E_3] = E_1 & [F_2,F_3] = -E_1 & [E_2,F_3] = F_1 & [F_2,E_3]=F_1 \\
\mbox{}[E_3,E_1] = -E_2 & [F_3,F_1] = E_2 & [E_3,F_1] = -F_2 & [F_3,E_1]=-F_2
\end{array}
\label{comm}
\end{equation}
All other commutators vanish. Abstractly, these relations define
the Lie algebra of the Lorentz group O(1,3), and the matrices
$E_i$ and $F_i$ defined by (\ref{ef}) correspond to a complex
representation of this algebra. 

\subsection{Another Representation}
Our goal in this section is to present an explicit
representation of the Lorentz algebra (\ref{comm}) in terms 
of $4 \times 4$ matrices
defined over a non-division algebra.
The non-division algebra will be discussed next.

\subsubsection{The Semi-Complex Number System}
We consider numbers of the form
\begin{equation}
x+ {\rm j}y,
\end{equation}
where $x$ and $y$ are real numbers, and ${\rm j}$ is a commuting
element satisfying the relation
\begin{equation}
 {\rm j}^2 = 1.
\end{equation}
The set of all such numbers forms a simple Clifford algebra,
which will be denoted by the symbol ${\bf D}$, and called the `semi-complex 
number system'. 
Addition, subtraction, and multiplication are defined in the obvious way:
\begin{eqnarray}
 (x_1+{\rm j}y_1) \pm (x_2+{\rm j}y_2) & = & 
 (x_1 \pm x_2) + {\rm j}(y_1 \pm y_2), \\
 (x_1+{\rm j}y_1) \cdot (x_2+{\rm j}y_2) & = & 
 (x_1 x_2 + y_1 y_2) + {\rm j} (x_1 y_2 + y_1 x_2).
\end{eqnarray}
Given any semi-complex number $w=x+{\rm j}y$, we define the 
`semi-complex conjugate' of $w$, written ${\overline w}$, to be 
\begin{equation}
 {\overline w} = x - {\rm j}y.
\end{equation}
It is easy to check the following; for any $w_1,w_2 \in {\bf D}$,
we have 
\begin{eqnarray}
{\overline{w_1+w_2}} & = & {\overline w_1} + {\overline w_2}, \\
{\overline{w_1\cdot w_2}} & = & {\overline w_1} \cdot {\overline w_2}.
\end{eqnarray}
We also have the identity
\begin{equation}
 {\overline w} \cdot w = x^2 - y^2
\end{equation}
for any semi-complex number $w=x+{\rm j}y$. Thus ${\overline w} \cdot w$
is always real, although unlike
the complex number system, it may take negative values.

At this point, it is convenient to define the `modulus squared'
of $w$, written $|w|^2$, as  
\begin{equation}
       |w|^2 = {\overline w} \cdot w.
\end{equation}
A nice consequence of these definitions is that for
any semi-complex numbers $w_1,w_2 \in {\bf D}$, we have
\begin{equation}
 |w_1 \cdot w_2|^2 = |w_1|^2 \cdot |w_2|^2.
\end{equation}
Now observe that if $|w|^2$ doesn't vanish, the quantity
\begin{equation}
w^{-1} = \frac{1}{|w|^2} \cdot {\overline{w}}
\end{equation}
is a well-defined unique inverse for $w$. So $w\in {\bf D}$ fails to
have an inverse if and only if $|w|^2 = x^2-y^2 =0$.
The semi-complex number system is therefore a non-division algebra.

\subsubsection{The Semi-Complex Unitary Groups}
Suppose $H$ is an $n \times n$ matrix defined over ${\bf D}$.
Then $H^{\dagger}$ will denote the $n\times n$ matrix
which is obtained by transposing $H$, and then conjugating 
each of the entries: $H^{\dagger} = \overline{H}^{T}$.
We say $H$ is Hermitian with respect to ${\bf D}$ if
$H^{\dagger} = H$, and anti-Hermitian if
$H^{\dagger} = -H$. 

Note that if $H$ is an $n \times n$ Hermitian matrix over ${\bf D}$,
then $U = e^{{\rm j}H}$ has the property 
\begin{equation}
           U^{\dagger}\cdot U = U \cdot U^{\dagger} = 1.
\label{un}
\end{equation}  
The set of all $n \times n$ matrices over ${\bf D}$ satisfying
the constraint (\ref{un}) forms a group, which we will
denote as U$(n,{\bf D})$, and call the `unitary group of
$n \times n$ semi-complex matrices'.
The `special unitary' subgroup SU($n,{\bf D}$) will be defined as all 
elements $U \in$U($n,{\bf D}$) satisfying the additional constraint
\begin{equation}
\det{U} = 1.
\end{equation}
Note that the semi-complex unitary groups we have defined above 
may be isomorphic to well known non-compact groups 
that are usually defined over the complex number field. 
For example, the semi-complex
group SU$(2,{\bf D})$ is isomorphic to the 
complex group SU(1,1) by virtue of the identification
\begin{equation}
\left(
\begin{array}{cc}
a_1 + {\rm i}a_2 & b_1 + {\rm i}b_2 \\
b_1 - {\rm i}b_2 & a_1 - {\rm i} a_2
\end{array} 
\right) \hspace{4mm}
\leftrightarrow \hspace{4mm}
\left(
\begin{array}{cc}
a_1 + {\rm j}b_1 & -a_2 + {\rm j}b_2 \\
a_2 + {\rm j}b_2 & a_1 - {\rm j} b_1 
\end{array}
\right),
\end{equation}
where the four real parameters $a_1,a_2,b_1$ and $b_2$ satisfy
the constraint $a_1^2+a_2^2-b_1^2-b_2^2=1$.

\subsubsection{A Semi-Complex Representation}
As promised, we will give an explicit representation of
the Lorentz algebra (\ref{comm}) in terms of matrices
defined over ${\bf D}$. First, we define three $2 \times 2$ matrices
${\bf \tau} = (\tau_1,\tau_2,\tau_3)$ by writing
\begin{equation}
\tau_1 = \left(
\begin{array}{cc}
 0 & 1 \\
1 & 0 
\end{array}
\right),
\hspace{5mm}
\tau_2 = \left(
\begin{array}{cc}
 0 & -{\rm j} \\
{\rm j} & 0 
\end{array}
\right),
\hspace{5mm}
\tau_3 = \left(
\begin{array}{cc}
 1 & 0 \\
 0 & -1 
\end{array}
\right).
\hspace{5mm}
\end{equation}
These matrices satisfy the following commutation relations:
\begin{equation}
[\tau_1,\tau_2]=2{\rm j} \tau_3 \hspace{6mm} 
[\tau_2,\tau_3]=2{\rm j} \tau_1 \hspace{6mm}
[\tau_3,\tau_1]=-2{\rm j} \tau_2   
\end{equation}
Now redefine the matrices $E_i$ and $F_i$, $i=1,2,3,$ by setting
\begin{equation}
 E_i = \frac{{\rm j}}{2}\left(
\begin{array}{cc}
 \tau_i & 0 \\
 0 & \tau_i
\end{array}
\right),
\hspace{6mm}
 F_i = \frac{1}{2}\left(
\begin{array}{cc}
 0 & \tau_i  \\
 -\tau_i & 0
\end{array}
\right), \hspace{5mm} i=1,2,3.
\label{semirep}
\end{equation}
The $4 \times 4$ 
semi-complex matrices $E_i$ and $F_i$
defined above are anti-Hermitian with respect to ${\bf D}$,
and satisfy the Lorentz algebra (\ref{comm}). We may therefore
introduce a semi-complex 4-component spinor $\Psi_{\bf D}$
transforming as follows under Lorentz transformations:
\begin{equation}
 \Psi_{\bf D} \rightarrow 
\exp{(\phi_1 E_1 - \theta_2 E_2 + \phi_3 E_3 + 
\theta_1 F_1 + \phi_2 F_2 + \theta_3 F_3)} \cdot  \Psi_{\bf D},
\label{transformII}
\end{equation}
which is evidently the semi-complex analogue of transformation 
(\ref{transformD}). Note that the transformation (\ref{transformII})
has the form $\Psi_{\bf D} \rightarrow U \cdot \Psi_{\bf D}$,
where $U \in$SU($4,{\bf D}$), since the generators $E_i$ and $F_i$
are traceless and anti-Hermitian with respect to ${\bf D}$. 
Thus the Lorentz group is a {\em subgroup} of the semi-complex
group SU($4,{\bf D}$). 

In the next section, we discuss a relation between the
complex Dirac spinor $\Psi_{\bf C}$, and the 4-component
semi-complex spinor $\Psi_{\bf D}$ defined above. 

\section{Equivalence of Spinor Representations}
\subsection{An Equivalence}
\label{isomorph}
Consider an {\em infinitesimal} Lorentz transformation of the
complex Dirac spinor,
\begin{equation}
\Psi_{\bf C} \rightarrow \exp{(\phi_1 E_1 - \theta_2 E_2 + \phi_3 E_3 + 
\theta_1 F_1 + \phi_2 F_2 + \theta_3 F_3)} \cdot  \Psi_{\bf C},
\label{transformsmall}
\end{equation}
where 
\begin{equation}
\Psi_{\bf C} =
\left(
\begin{array}{c}
x_1+{\rm i}y_1 \\
x_2+{\rm i}y_2 \\
x_3+{\rm i}y_3 \\
x_4+{\rm i}y_4
\end{array}
\label{cd}
\right).
\end{equation}
In terms of the eight real variables
$x_i$ and $y_i$, $i=1,2,3,4$, an infinitesimal
transformation of the form (\ref{transformsmall}) 
is equivalent to the following eight real transformations:
\begin{eqnarray}
 x_1 & \rightarrow & x_1 + \frac{\phi_1}{2} \cdot x_2
                         + \frac{\theta_2}{2} \cdot x_2 
                         + \frac{\phi_3}{2} \cdot x_1 
                         - \frac{\theta_1}{2} \cdot y_2
                         + \frac{\phi_2}{2} \cdot y_2 
                         - \frac{\theta_3}{2} \cdot y_1 \nonumber \\
 x_2 & \rightarrow & x_2 + \frac{\phi_1}{2} \cdot x_1
                         - \frac{\theta_2}{2} \cdot x_1 
                         - \frac{\phi_3}{2} \cdot x_2 
                         - \frac{\theta_1}{2} \cdot y_1
                         - \frac{\phi_2}{2} \cdot y_1 
                         + \frac{\theta_3}{2} \cdot y_2 \nonumber \\
 x_3 & \rightarrow & x_3 - \frac{\phi_1}{2} \cdot x_4
                         + \frac{\theta_2}{2} \cdot x_4 
                         - \frac{\phi_3}{2} \cdot x_3
                         - \frac{\theta_1}{2} \cdot y_4
                         - \frac{\phi_2}{2} \cdot y_4
                         - \frac{\theta_3}{2} \cdot y_3 \nonumber \\
 x_4 & \rightarrow & x_4 - \frac{\phi_1}{2} \cdot x_3
                         - \frac{\theta_2}{2} \cdot x_3 
                         + \frac{\phi_3}{2} \cdot x_4
                         - \frac{\theta_1}{2} \cdot y_3
                         + \frac{\phi_2}{2} \cdot y_3
                         + \frac{\theta_3}{2} \cdot y_4 \nonumber \\
 y_1 & \rightarrow & y_1 + \frac{\phi_1}{2} \cdot y_2
                         + \frac{\theta_2}{2} \cdot y_2 
                         + \frac{\phi_3}{2} \cdot y_1
                         + \frac{\theta_1}{2} \cdot x_2
                         - \frac{\phi_2}{2} \cdot x_2
                         + \frac{\theta_3}{2} \cdot x_1 \nonumber \\
 y_2 & \rightarrow & y_2 + \frac{\phi_1}{2} \cdot y_1
                         - \frac{\theta_2}{2} \cdot y_1 
                         - \frac{\phi_3}{2} \cdot y_2
                         + \frac{\theta_1}{2} \cdot x_1
                         + \frac{\phi_2}{2} \cdot x_1
                         - \frac{\theta_3}{2} \cdot x_2 \nonumber \\
 y_3 & \rightarrow & y_3 - \frac{\phi_1}{2} \cdot y_4
                         + \frac{\theta_2}{2} \cdot y_4 
                         - \frac{\phi_3}{2} \cdot y_3
                         + \frac{\theta_1}{2} \cdot x_4
                         + \frac{\phi_2}{2} \cdot x_4
                         + \frac{\theta_3}{2} \cdot x_3 \nonumber \\
 y_4 & \rightarrow & y_4 - \frac{\phi_1}{2} \cdot y_3
                         - \frac{\theta_2}{2} \cdot y_3 
                         + \frac{\phi_3}{2} \cdot y_4
                         + \frac{\theta_1}{2} \cdot x_3
                         - \frac{\phi_2}{2} \cdot x_3
                         - \frac{\theta_3}{2} \cdot x_4 
\label{smallx}
\end{eqnarray}
where we have used the representation specified by (\ref{ef}).
Now consider the corresponding
infinitesimal Lorentz transformation of the semi-complex spinor
$\Psi_{\bf D}$,
\begin{equation}
\Psi_{\bf D} \rightarrow \exp{(\phi_1 E_1 - \theta_2 E_2 + \phi_3 E_3 + 
\theta_1 F_1 + \phi_2 F_2 + \theta_3 F_3)} \cdot  \Psi_{\bf D},
\label{transformsmallII}
\end{equation}
where 
\begin{equation}
\Psi_{\bf D} =
\left(
\begin{array}{c}
a_1+{\rm j}b_1 \\
a_2+{\rm j}b_2 \\
a_3+{\rm j}b_3 \\
a_4+{\rm j}b_4
\end{array}
\right).
\end{equation}
In terms of the eight real variables $a_i$ and $b_i$, $i=1,2,3,4$,
this infinitesimal transformation is equivalent to the following
eight real transformations:
\begin{eqnarray}
 a_1 & \rightarrow & a_1 + \frac{\phi_1}{2} \cdot b_2
                         + \frac{\theta_2}{2} \cdot a_2 
                         + \frac{\phi_3}{2} \cdot b_1 
                         + \frac{\theta_1}{2} \cdot a_4
                         - \frac{\phi_2}{2} \cdot b_4 
                         + \frac{\theta_3}{2} \cdot a_3 \nonumber \\
 a_2 & \rightarrow & a_2 + \frac{\phi_1}{2} \cdot b_1
                         - \frac{\theta_2}{2} \cdot a_1 
                         - \frac{\phi_3}{2} \cdot b_2 
                         + \frac{\theta_1}{2} \cdot a_3
                         + \frac{\phi_2}{2} \cdot b_3 
                         - \frac{\theta_3}{2} \cdot a_4 \nonumber \\
 a_3 & \rightarrow & a_3 + \frac{\phi_1}{2} \cdot b_4
                         + \frac{\theta_2}{2} \cdot a_4 
                         + \frac{\phi_3}{2} \cdot b_3
                         - \frac{\theta_1}{2} \cdot a_2
                         + \frac{\phi_2}{2} \cdot b_2
                         - \frac{\theta_3}{2} \cdot a_1 \nonumber \\
 a_4 & \rightarrow & a_4 + \frac{\phi_1}{2} \cdot b_3
                         - \frac{\theta_2}{2} \cdot a_3 
                         - \frac{\phi_3}{2} \cdot b_4
                         - \frac{\theta_1}{2} \cdot a_1
                         - \frac{\phi_2}{2} \cdot b_1
                         + \frac{\theta_3}{2} \cdot a_2 \nonumber \\
 b_1 & \rightarrow & b_1 + \frac{\phi_1}{2} \cdot a_2
                         + \frac{\theta_2}{2} \cdot b_2 
                         + \frac{\phi_3}{2} \cdot a_1
                         + \frac{\theta_1}{2} \cdot b_4
                         - \frac{\phi_2}{2} \cdot a_4
                         + \frac{\theta_3}{2} \cdot b_3 \nonumber \\
 b_2 & \rightarrow & b_2 + \frac{\phi_1}{2} \cdot a_1
                         - \frac{\theta_2}{2} \cdot b_1 
                         - \frac{\phi_3}{2} \cdot a_2
                         + \frac{\theta_1}{2} \cdot b_3
                         + \frac{\phi_2}{2} \cdot a_3
                         - \frac{\theta_3}{2} \cdot b_4 \nonumber \\
 b_3 & \rightarrow & b_3 + \frac{\phi_1}{2} \cdot a_4
                         + \frac{\theta_2}{2} \cdot b_4 
                         + \frac{\phi_3}{2} \cdot a_3
                         - \frac{\theta_1}{2} \cdot b_2
                         + \frac{\phi_2}{2} \cdot a_2
                         - \frac{\theta_3}{2} \cdot b_1 \nonumber \\
 b_4 & \rightarrow & b_4 + \frac{\phi_1}{2} \cdot a_3
                         - \frac{\theta_2}{2} \cdot b_3 
                         - \frac{\phi_3}{2} \cdot a_4
                         - \frac{\theta_1}{2} \cdot b_1
                         - \frac{\phi_2}{2} \cdot a_1
                         + \frac{\theta_3}{2} \cdot b_2 
\label{smalla}
\end{eqnarray}
where we have used the representation specified by the semi-complex
matrices (\ref{semirep}).

It is now straightforward to check that the infinitesimal 
transformations (\ref{smallx}) and (\ref{smalla}) for the
complex and semi-complex spinors respectively are {\em equivalent}
if we make the following identifications\footnote{The factor of
$1/\sqrt{2}$ is arbitrary, and introduced for later convenience.}:
\begin{equation}
\begin{array}{cccc}
a_1 \leftrightarrow \frac{1}{\sqrt{2}}(y_1+y_3) \hspace{4mm} & 
a_2 \leftrightarrow \frac{1}{\sqrt{2}}(y_2 + y_4) \hspace{4mm}&
a_3 \leftrightarrow \frac{1}{\sqrt{2}}(x_1 + x_3) \hspace{4mm} &
a_4 \leftrightarrow \frac{1}{\sqrt{2}}(x_2 + x_4)  \\
b_1 \leftrightarrow \frac{1}{\sqrt{2}}(y_1-y_3) \hspace{4mm} &
b_2 \leftrightarrow \frac{1}{\sqrt{2}}(y_2-y_4) \hspace{4mm}  &
b_3 \leftrightarrow \frac{1}{\sqrt{2}}(x_1 - x_3) \hspace{4mm}&
b_4 \leftrightarrow \frac{1}{\sqrt{2}}(x_2 - x_4) 
\end{array} 
\end{equation}
In particular, we have the identification
\begin{equation}
(I) \hspace{3mm} \Psi_{\bf C} =
\left(
\begin{array}{c}
x_1+{\rm i}y_1 \\
x_2+{\rm i}y_2 \\
x_3+{\rm i}y_3 \\
x_4+{\rm i}y_4
\end{array}
\right) \hspace{5mm} \leftrightarrow \hspace{5mm}
\Psi_{\bf D} =
\frac{1}{\sqrt{2}}\left(
\begin{array}{c}
(y_1+y_3)  +{\rm j}(y_1-y_3)  \\
(y_2 + y_4) +{\rm j}(y_2-y_4)  \\
(x_1 + x_3) +{\rm j}(x_1 - x_3) \\
(x_2 + x_4)  +{\rm j}(x_2 - x_4)
\end{array}
\right),
\label{iso}
\end{equation}   
which establishes an exact equivalence between a complex
Lorentz transformation [Eqn(\ref{transformD})] 
acting on the Dirac 4-spinor
$\Psi_{\bf C}$, and the corresponding
Lorentz transformation [Eqn(\ref{transformII})]
acting on a semi-complex 4-spinor $\Psi_{\bf D}$.

It turns out that
the equivalence specified by the identification (\ref{iso})
is not unique. There are additional identifications that
render the complex and semi-complex Lorentz transformations
equivalent, and we list three more below:
\begin{equation}
(II) \hspace{3mm} \left(
\begin{array}{c}
x_1+{\rm i}y_1 \\
x_2+{\rm i}y_2 \\
x_3+{\rm i}y_3 \\
x_4+{\rm i}y_4
\end{array}
\right) \hspace{3mm} \leftrightarrow \hspace{3mm}
\frac{1}{\sqrt{2}}\left(
\begin{array}{c}
-(y_2+y_4)  +{\rm j}(y_2-y_4)  \\
(y_1 + y_3) -{\rm j}(y_1-y_3)  \\
(x_2 + x_4) -{\rm j}(x_2 - x_4) \\
-(x_1 + x_3)  +{\rm j}(x_1 - x_3)
\end{array}
\right),
\end{equation}
\begin{equation}
(III) \hspace{3mm} 
\left(
\begin{array}{c}
x_1+{\rm i}y_1 \\
x_2+{\rm i}y_2 \\
x_3+{\rm i}y_3 \\
x_4+{\rm i}y_4
\end{array}
\right) \hspace{3mm} \leftrightarrow \hspace{3mm}
\frac{1}{\sqrt{2}}\left(
\begin{array}{c}
-(x_1+x_3)  -{\rm j}(x_1-x_3)  \\
-(x_2 + x_4) -{\rm j}(x_2-x_4)  \\
(y_1 + y_3)  +{\rm j}(y_1 - y_3) \\
(y_2 + y_4)  +{\rm j}(y_2 - y_4)
\end{array}
\right),
\end{equation} 
and
\begin{equation}
(IV) \hspace{3mm} 
\left(
\begin{array}{c}
x_1+{\rm i}y_1 \\
x_2+{\rm i}y_2 \\
x_3+{\rm i}y_3 \\
x_4+{\rm i}y_4
\end{array}
\right) \hspace{3mm} \leftrightarrow \hspace{3mm}
\frac{1}{\sqrt{2}}\left(
\begin{array}{c}
-(x_2+x_4)  +{\rm j}(x_2-x_4)  \\
(x_1 + x_3) -{\rm j}(x_1-x_3)  \\
-(y_2 + y_4) +{\rm j}(y_2 - y_4) \\
(y_1 + y_3)  -{\rm j}(y_1 - y_3)
\end{array}
\right).
\end{equation}
Four more identifications may be obtained
by a simple `reflection' procedure; simply multiply
each semi-complex spinor appearing in identifications
(I),(II),(III) and (IV) above by the semi-complex 
variable ${\rm j}$. This has the effect of interchanging
the `real' and `imaginary' parts of each semi-complex component
in the spinor.  

\subsection{Parity}
Under parity, the Dirac 4-spinor $\Psi_{\bf C}$ transforms as follows
\cite{ryder}:
\begin{equation}
 \Psi_{\bf C} \rightarrow
\left(
\begin{array}{cc}
 0 & \mbox{{\bf 1}}_{2 \times 2} \\
 \mbox{{\bf 1}}_{2 \times 2} & 0
\end{array}
\right)
\cdot
\Psi_{\bf C},
\end{equation}
or, in terms of the eight real components $x_i$ and $y_i$,
$i=1,2,3,4$, of the Dirac 4-spinor
$\Psi_{\bf C}$ specified by (\ref{cd}), we have
\begin{equation}
\begin{array}{cccc}
 x_1 \rightarrow x_3 \hspace{5mm}  &
x_2 \rightarrow x_4 \hspace{5mm}  &
x_3 \rightarrow x_1 \hspace{5mm}  &
x_4 \rightarrow x_2 \\
 y_1 \rightarrow y_3 \hspace{5mm}  &
y_2 \rightarrow y_4 \hspace{5mm}  &
y_3 \rightarrow y_1 \hspace{5mm}  &
y_4 \rightarrow y_2 
\end{array}
\end{equation}
According to  the identifications (I),(II),(III) and (IV)
of Section \ref{isomorph},
a parity transformation on $\Psi_{\bf C}$ corresponds 
to conjugating $\Psi_{\bf D}$. Thus,
$\Psi_{\bf D} \rightarrow \Psi_{\bf D}^{\ast}$ under parity\footnote{
$\Psi_{\bf D}^{\ast}$ denotes taking the semi-complex conjugate
of each element in $\Psi_{\bf D}$.} for these identifications.
The `reflected' forms of these identifications 
induces the transformation $\Psi_{\bf D} \rightarrow {\rm j}
\Psi_{\bf D}^{\ast}$ under parity. Semi-complex conjugation 
is therefore closely related to the parity symmetry.

\section{The Automorphism Algebra of the Dirac Spinor}
\label{auto} 
The existence of distinct equivalences between the complex 
and semi-complex spinors permits one to construct automorphisms
of the complex Dirac spinor that leave the transformation
properties of its eight real components intact under
Lorentz transformations.

In order to investigate the algebra underlying
the set of all possible automorphisms,
it is convenient to change our current basis to
the so-called `standard representation' of the 
Lorentz group \cite{ryder}. The Dirac 4-spinor
$\Psi^{SR}_{\bf C}$ 
in the standard representation
is related to the original 4-spinor $\Psi_{\bf C}$
according to the relation
\begin{equation}
\Psi^{SR}_{\bf C}
=
\frac{1}{\sqrt{2}}
\left(
\begin{array}{cc}
1 & 1 \\
1 & -1 
\end{array}
\right) \cdot \Psi_{\bf C}.
\end{equation}
The identifications (I)-(IV) stated in Section \ref{isomorph}
are now equivalent to the following identifications:
\begin{equation}
(I)' \hspace{7mm} 
\Psi_{\bf D} =
\left(
\begin{array}{c}
a_1 +{\rm j}b_1  \\
a_2 +{\rm j}b_2  \\
a_3 +{\rm j}b_3 \\
a_4 +{\rm j}b_4
\end{array}
\right) \hspace{5mm} \leftrightarrow \hspace{5mm}
\Psi^{SR}_{\bf C} =
\left(
\begin{array}{c}
a_3+{\rm i}a_1 \\
a_4+{\rm i}a_2 \\
b_3+{\rm i}b_1 \\
b_4+{\rm i}b_2
\end{array}
\right) 
\end{equation}
\begin{equation}
(II)' \hspace{7mm} 
\Psi_{\bf D} =
\left(
\begin{array}{c}
a_1 +{\rm j}b_1  \\
a_2 +{\rm j}b_2  \\
a_3 +{\rm j}b_3 \\
a_4 +{\rm j}b_4
\end{array}
\right) \hspace{5mm} \leftrightarrow \hspace{5mm}
\Psi^{SR}_{\bf C} =
\left(
\begin{array}{c}
-a_4+{\rm i}a_2 \\
a_3-{\rm i}a_1 \\
b_4-{\rm i}b_2 \\
-b_3+{\rm i}b_1
\end{array}
\right) 
\end{equation}
\begin{equation}
(III)' \hspace{7mm} 
\Psi_{\bf D} =
\left(
\begin{array}{c}
a_1 +{\rm j}b_1  \\
a_2 +{\rm j}b_2  \\
a_3 +{\rm j}b_3 \\
a_4 +{\rm j}b_4
\end{array}
\right) \hspace{5mm} \leftrightarrow \hspace{5mm}
\Psi^{SR}_{\bf C} =
\left(
\begin{array}{c}
-a_1+{\rm i}a_3 \\
-a_2+{\rm i}a_4 \\
-b_1+{\rm i}b_3 \\
-b_2+{\rm i}b_4
\end{array}
\right) 
\end{equation} 
\begin{equation}
(IV)' \hspace{7mm} 
\Psi_{\bf D} =
\left(
\begin{array}{c}
a_1 +{\rm j}b_1  \\
a_2 +{\rm j}b_2  \\
a_3 +{\rm j}b_3 \\
a_4 +{\rm j}b_4
\end{array}
\right) \hspace{5mm} \leftrightarrow \hspace{5mm}
\Psi^{SR}_{\bf C} =
\left(
\begin{array}{c}
a_2+{\rm i}a_4 \\
-a_1-{\rm i}a_3 \\
-b_2-{\rm i}b_4 \\
b_1+{\rm i}b_3
\end{array}
\right).
\end{equation} 
In addition, we have four
more which correspond to the `reflected' form
of the above identifications, and are obtained by interchanging
the real and imaginary parts of the semi-complex components of
$\Psi_{\bf D}$:
\begin{equation}
(V)' \hspace{7mm} 
\Psi_{\bf D} =
\left(
\begin{array}{c}
a_1 +{\rm j}b_1  \\
a_2 +{\rm j}b_2  \\
a_3 +{\rm j}b_3 \\
a_4 +{\rm j}b_4
\end{array}
\right) \hspace{5mm} \leftrightarrow \hspace{5mm}
\Psi^{SR}_{\bf C} =
\left(
\begin{array}{c}
b_3+{\rm i}b_1 \\
b_4+{\rm i}b_2 \\
a_3+{\rm i}a_1 \\
a_4+{\rm i}a_2
\end{array}
\right) 
\end{equation}
\begin{equation}
(VI)' \hspace{7mm} 
\Psi_{\bf D} =
\left(
\begin{array}{c}
a_1 +{\rm j}b_1  \\
a_2 +{\rm j}b_2  \\
a_3 +{\rm j}b_3 \\
a_4 +{\rm j}b_4
\end{array}
\right) \hspace{5mm} \leftrightarrow \hspace{5mm}
\Psi^{SR}_{\bf C} =
\left(
\begin{array}{c}
-b_4+{\rm i}b_2 \\
b_3-{\rm i}b_1 \\
a_4-{\rm i}a_2 \\
-a_3+{\rm i}a_1
\end{array}
\right) 
\end{equation}
\begin{equation}
(VII)' \hspace{7mm} 
\Psi_{\bf D} =
\left(
\begin{array}{c}
a_1 +{\rm j}b_1  \\
a_2 +{\rm j}b_2  \\
a_3 +{\rm j}b_3 \\
a_4 +{\rm j}b_4
\end{array}
\right) \hspace{5mm} \leftrightarrow \hspace{5mm}
\Psi^{SR}_{\bf C} =
\left(
\begin{array}{c}
-b_1+{\rm i}b_3 \\
-b_2+{\rm i}b_4 \\
-a_1+{\rm i}a_3 \\
-a_2+{\rm i}a_4
\end{array}
\right) 
\end{equation} 
\begin{equation}
(VIII)' \hspace{7mm} 
\Psi_{\bf D} =
\left(
\begin{array}{c}
a_1 +{\rm j}b_1  \\
a_2 +{\rm j}b_2  \\
a_3 +{\rm j}b_3 \\
a_4 +{\rm j}b_4
\end{array}
\right) \hspace{5mm} \leftrightarrow \hspace{5mm}
\Psi^{SR}_{\bf C} =
\left(
\begin{array}{c}
b_2+{\rm i}b_4 \\
-b_1-{\rm i}b_3 \\
-a_2-{\rm i}a_4 \\
a_1+{\rm i}a_3
\end{array}
\right).
\end{equation} 
Recall what these identifications mean; namely, under any
given semi-complex Lorentz transformation [Eqn(\ref{transformII})] of
$\Psi_{\bf D}$, the eight real components $a_i$ and $b_i$
($i=1,2,3,4$) transform in exactly the same way as the
eight real components $a_i$ and $b_i$ that appear in
the (eight) complex spinors
$\Psi^{SR}_{\bf C}$ listed
above, after being acted on by the 
corresponding complex Lorentz transformation\footnote{
We assume the $E_i$'s and $F_i$'s are now in the standard representation.} 
[Eqn(\ref{transformD})].

We now define an operator $\rho_{II}$ which takes the complex spinor
$\Psi^{SR}_{\bf C}$ in the identification (I)' above 
and maps it to the complex spinor $\Psi^{SR}_{\bf C}$
in the identification (II)'. Thus $\rho_{II}$ is defined by
\begin{equation}
\rho_{II}\cdot
\left(
\begin{array}{c}
x_1+{\rm i}y_1 \\
x_2+{\rm i}y_2 \\
x_3+{\rm i}y_3 \\
x_4+{\rm i}y_4
\end{array}
\right)
 =
\left(
\begin{array}{c}
-x_2+{\rm i}y_2 \\
x_1-{\rm i}y_1 \\
x_4-{\rm i}y_4 \\
-x_3+{\rm i}y_3
\end{array}
\right),
\end{equation}
for any real variables $x_i$ and $y_i$.
Similarly, we may construct the operators
$\rho_{III}, \rho_{IV}, \dots , \rho_{VIII}$,
whose explicit form we omit for brevity.

If we let 
${\cal V}(\Psi^{SR}_{\bf C})$ denote the eight-dimensional 
vector space 
formed by all {\em real} linear combinations of 
complex 4-spinors, then the linear map $\rho_{II}$,
for example, is
an automorphism of ${\cal V}(\Psi^{SR}_{\bf C})$.
In particular, the transformation properties
of the eight real components of $\Psi^{SR}_{\bf C}$
under a Lorentz transformation is identical to the 
transformation properties of the transformed spinor
$\rho_{II}(\Psi^{SR}_{\bf C})$ under the same Lorentz transformation.
One can show that the set of eight operators
\begin{equation}
 \{1,\rho_{II},\rho_{III},\dots,\rho_{VIII}\} 
\end{equation}
generate an eight dimensional closed algebra with respect to
the real numbers.. 
The subset $\{1,\rho_{II},\rho_{III},\rho_{IV}\}$, for
example, generates the algebra of quaternions. 

One may also consider all commutators of the seven elements
$\rho_{II},\rho_{III},\dots,\rho_{VIII}$. These turn out to
generate a Lie algebra that is isomorphic to 
$\mbox{SU(2)}\times\mbox{SU(2)}\times\mbox{U(1)}$.
The $\mbox{SU(2)}\times\mbox{SU(2)}$ part 
is a Lorentz symmetry. The U(1) factor is
intriguing.

\section{Discussion} 
In this work, we constructed a representation of
the six-dimensional Lorentz group in terms of 
$4 \times 4$ generating matrices
defined over a non-division algebra.
This algebra was referred to as the `semi-complex' number system.

The transformation properties
of the corresponding `semi-complex 4-spinor' 
was shown to be equivalent to the transformation 
properties of the usual complex Dirac spinor, after
making an appropriate identification of components. 
The non-uniqueness
of this identification led to an automorphism algebra 
defined on the vector space of Dirac spinors.
These automorphisms have the property of preserving the
transformation properties of the eight-real components
of a 4-spinor in any given Lorentz frame. Properties
of this algebra were studied.

It is interesting to note that the semi-complex representation
of the Lorentz group turns out to be a subgroup of SU(4,{\bf D}).
This group is fifteen dimensional, and so a reasonable guess
is that 
it is isomorphic to the conformal group SU(2,2), since we know  
SU$(2,{\bf D}) \cong $SU(1,1).  However, this is not the case.
By noting that the generators of  SU(4,{\bf D}) are $4\times 4$
anti-Hermitian matrices with respect to ${\bf D}$,
and that the semi-complex variable ${\rm j}$ may
be mapped to the element $1$ or $-1$, one can show that the Lie algebra
of SU(4,{\bf D}) is isomorphic to SL(4,{\bf R}). But we also know
SL(4,{\bf R})$\cong$SO(3,3) \cite{group}. Thus the semi-complex
Lorentz group is naturally embedded in a group with SO(3,3)
symmetry, rather than the conformal symmetry SO(2,4).

Thus, from the viewpoint of naturally
embedding the Lorentz symmetry into some larger group,
the semi-complex and complex representations stand apart.
We leave the physics of SU(4,{\bf D}) as an intriguing 
topic yet to be studied.

 \medskip

\begin{large}{\bf Acknowledgment}
\end{large}
I would like to thank the British Council for support
during the early stages of this work.

\end{document}